\documentstyle[aps,eqsecnum]{revtex}
\begin{document}
\title{An overview of canonical quantum gravity}
\author{Jorge Pullin}
\address{Center for Gravitational Physics and Geometry,
Department of Physics, The Pennsylvania State University,\\
104 Davey Lab, University Park, PA 16802}
\date{June 30, 1998}
\maketitle

\begin{abstract}
This is a summary of a talk delivered at the workshop ``Quantum
gravity in the Southern Cone II''. We present a very brief review of
current results on canonical quantization of general relativity using
Ashtekar's variables and loop quantization.
\end{abstract}
\vspace{-5cm} 
\begin{flushright}
\baselineskip=15pt
CGPG-98/7-1  \\
gr-qc/9806119\\
\end{flushright}
\vspace{3cm}

\section{Introduction}

Slightly over ten years ago, a new set of paths were opened in the
quest to try to apply the rules of quantum mechanics to general
relativity. Several researchers have contributed to this program,
developing various points and lines of thought. The common threads of
this work have to do with the use of a new set of variables to
describe the gravitational field, introduced by Ashtekar \cite{As86},
that make the canonical general relativity resemble a Yang--Mills
theory.

At first some people might find the premise of the whole program
questionable. After all, general relativity is known to be
power-counting-non-renormalizable; string theory suggests extra
structures are needed for a consistent theory; moreover, even quantum
field theory in curved space-times encounters serious difficulties.
Why waste effort attempting a program that appears surrounded by doom?
It should be realized, however, that most of the above statements are
made within the context of perturbation theory. Most of the detailed
calculations also introduce artificial background structures (for
instance, a Minkowski background in usual perturbative calculations)
in the theory that conflict with the basic symmetry of general
relativity: diffeomorphism invariance. In fact, if one takes seriously
the idea that diffeomorphism invariance is a fundamental symmetry of
nature, one might be led to believe that the proposed program is quite
sensible. Diffeomorphism invariance is usually associated with
non-perturbative, discrete objects and concepts. These are quite
distinct from those normally used in quantum field theories. It might
be the case that by attempting to impose diffeomorphism invariance, we
will be forced to use techniques and ideas that are quite novel. In
fact, the emergence of discrete structures and of fundamental
dynamical length scales (like the Planck length), suggests that the
ultraviolet problem of quantum field theories might be
controlled. These ideas have been argued back and forth in generic
terms for many years. However, the approach we will describe here
actually puts them into practice and reaches conclusions about the
points raised in a rather concrete fashion. In the end, the proof of
the validity of what is being attempted will be given by the emergence
of concrete physical predictions, hopefully testable, from the
theory. We will show that although we still do not have complete
control of the theory, some physical predictions are already emerging.

This paper will be a very quick and succinct review. It will not
attempt to make a detailed case for the various aspects mentioned. It
will be skimpy on explicit formulae. It will be incomplete in
referring to previous work. It just attempts to be a quickly readable
manuscript to introduce someone to some of the ideas in the subject.
For a more comprehensive review see the article by Rovelli in 
Living Reviews \cite{Roliving}.

\section{Canonical gravity}

Let us start with the basic setting. We will attempt a canonical
quantization of gravity. This will require setting the theory in a
Hamiltonian form. This has been studied by many authors (see
\cite{Asbook} for references). The idea is that one foliates
space-time into space and time and considers as fundamental canonical
variables the three metric $q^{ab}$ and as canonically conjugate
momentum a quantity that is closely related to the extrinsic curvature
$K_{ab}$. The time-time and the space-time portions of the space-time
metric (known as the lapse and shift vector) appear as Lagrange
multipliers in the action, which means that the theory has
constraints. In total there are four constraints, that structure
themselves into a vector and a scalar. These constraints are the
imprint in the canonical theory of the diffeomorphism invariance of
the four-dimensional theory. They also contain the dynamics of the
theory, since the Hamiltonian identically vanishes. This is not
surprising, it is the way in which the canonical formalism tells us
that the split into space and time that we perform is a fiduciary
one. If one attempts to quantize this theory one starts by choosing a
polarization for the wavefunctions (usually functions of the three
metric) and one has to implement the constraints as operator
equations. These will assure that the wavefunctions embody the
symmetries of the theory. The diffeomorphism constraint has a
geometrical interpretation, demanding that the wavefunctions be
functions of the ``three-geometry'' and not of the three-metric, that
is, that they be invariant under diffeomorphisms of the three
manifold. The Hamiltonian constraint does not admit a simple geometric
interpretation and should be implemented as an operatorial
equation. Unfortunately, it is a complicated non-polynomial function
of the basic variables and little progress had been made towards
realizing it as a quantum operator ever since De Witt considered the
problem in the 60's. Let us recall that in this context regularization
is a highly non-trivial process, since most common regulators used in
quantum field theory violate diffeomorphism invariance. Even if we
ignore these technical details, the resulting theory appears as very
difficult to interpret. The theory has no explicit dynamics, one is in
the ``frozen formalism''. Wavefunctions are annihilated by the
constraints and observable quantities commute with the
constraints. Observables are better described, as Kucha\v{r}
emphasizes, as ``perennials". The expectation is that in physical
situations some of the variables of the theory will play the role of
``time'' and in terms of them one would be able to define a ``true"
dynamics in a relational way, and a non-vanishing
Hamiltonian. Unfortunately, this has never been fully implemented in
practice and in fact several model examples show that the idea can
quickly run into trouble \cite{torre}. This is the content of the
``problem of time'' in canonical quantization, which has been
discussed extensively in the literature \cite{kuchartime}.

About ten years ago, Ashtekar \cite{As86} proposed the use of a new
set of variables to describe canonical gravity. These new variables
had the advantage that they made the theory resemble a Yang--Mills
theory. This opened the possibility of importing techniques from the
Yang--Mills context to the gravitational one. The easiest way to
introduce the new variables (as pointed out by Barbero \cite{Bar}) is
to go through an intermediate step, reformulating ordinary canonical
gravity in terms of triads instead of metrics. Canonical gravity has
been described in terms of triads in the past (see for instance
\cite{He}). The canonical variables are three frame fields
$\tilde{E}^a_i$ and the conjugate variables $K_a^i$ are again closely
related to the extrinsic curvature. We use a tilde to denote density
weights. The theory has the usual diffeomorphism and Hamiltonian
constraints plus three additional constraints that state that the
theory is invariant under triad rotations. The Hamiltonian constraint
is, as in the usual metric variables, non-polynomial. If in this
theory we now perform a canonical transformation defining a new
variable $A_a^i = \Gamma_a^i+\beta K_a^i$, where $\Gamma_a^i$ is the
metric-compatible spin connection, and $\beta$ is a parameter, and we
re-scale the triads by $1/\beta$, the constraints of the theory read,
\begin{eqnarray}
D_a \tilde{E}^a_i &=& 0\\
\tilde{E}^a_i F_{ab}^i &=&0\\
\epsilon_{ijk} \tilde{E}^a_i\tilde{E}^b_j F_{ab}^k +
2{(1+\beta^2)\over \beta} \tilde{E}^a_i\tilde{E}^b_j
(A_a^i-\Gamma_a^i)
(A_b^j-\Gamma_b^j) &=&0.
\end{eqnarray}
As we see, the first constraint has exactly the form of a Yang--Mills
type Gauss law ($D_a$ is the derivative defined by the connection
$A_a^i$). The second constraint (the diffeomorphism or momentum
constraint) states that the Poynting vector of the field vanishes. The
last constraint (also called the Hamiltonian constraint, or, when
promoted to a quantum operator, the Wheeler--DeWitt equation) is
non-polynomial due to the presence of the $\Gamma_a^i$ terms. If one
were to choose the parameter $\beta$ equal to the imaginary unit, the
last term disappears and the constraints become polynomial. This was
the original way in which the Ashtekar variables were introduced. The
polynomiality of the constraint led quickly to a quantum
representation and the discovery of solutions, for the first time
ever, of the Wheeler--DeWitt equation \cite{JaSm}. Nowaday, the
variables is slightly different. Thiemann \cite{Th1-6} has shown that
if one divides the Hamiltonian constraint by the square root of the
determinant of the three metric, both pieces of the constraint can be
made polynomial in terms of Poisson brackets of the connection and the
volume of the space. This has led to a series of articles in which he
has explored the quantum implementation of the constraints, that we
will comment upon later. Therefore, there is no need, from the
polynomiality standpoint, to set the value of $\beta$ to the imaginary
unit. This has additional advantages. If $\beta$ is complex one is
dealing with a real theory described in terms of complex
variables. After quantization, reality has to be recovered. This led
to lengthy discussions about how to implement the ``reality
conditions'' that assured that the theory was a real one. If one takes
real values for $\beta$, all variables in the theory are real. The
resulting theory has a more complicated ---yet polynomial---
Hamiltonian constraint than the one with $\beta=i$, but it still
retains the Gauss law and Poynting vector conditions (which are
$\beta$-independent). We can still think of the phase space of the
theory as a sub-manifold of the phase space of an $SU(2)$ Yang--Mills
theory, and apply several of the techniques we will discuss. We will
therefore adopt the viewpoint from now on that the variables used are
real.

\section{Quantization and loops}

To proceed with the canonical quantization of the theory it is
customary to pick a polarization where wavefunctions are functions of
the connection $A_a^i$, this polarization would be the usual one used
in a Yang--Mills context. The Gauss law requires that the wavefunction
be $SU(2)$ (gauge) invariant. The diffeomorphism constraint requires
that the wavefunctions be diffeomorphism invariant. In this context it
turns out to be useful to consider a particular set of wavefunctions,
parameterized by loops, called Wilson loops. These are constructed
considering the trace of the holonomy of the connection $A_a^i$ around
a closed loop $\gamma$. These objects are gauge invariant and
therefore solve automatically the Gauss law. In fact, there are a
``basis'' of gauge invariant functions in the sense that one can
reconstruct all gauge invariant information in the connection from
them \cite{Gi}, and therefore any function of the connection as
well. These reasons favored their use for quite some time in the
Yang--Mills context. Unfortunately, the Wilson loops $W_\gamma(A)$ are
not independent functions. Wilson loops based on different loops
satisfy identities (called Mandelstam identities \cite{GaTr86}) that
constrain their values. Therefore they are not really a ``basis'' but
an over-complete basis. We will address this point later on, at the
moment we will loosely refer to Wilson loops as a basis.

If one expands a given wavefunction $\psi(A)$ in terms of the Wilson
loop basis, 
\begin{equation}
\psi(\gamma) = \int DA \psi(A) W_\gamma(A) \label{looptransf})
\end{equation}
the coefficients in the expansion will be functions of the loops
$\psi(\gamma)$ and will contain all the information $\psi(A)$
contained. One can view these coefficients as wavefunctions in a new
representation of the theory called the loop representation. As an analogy,
one can consider the position and momentum representations in ordinary
quantum mechanics and the Fourier transform between them. In the loop 
representation, wavefunctions are functions of loops, the Gauss law is
already solved, and operators will have to be geometrical in nature,
acting upon the loops. This representation had been studied by Gambini
and Trias in the Yang--Mills context in the early eighties with some
success, including lattice gauge theory calculations \cite{GaTr86}. An
added bonus in the gravitational context, first noted by Rovelli and
Smolin \cite{RoSm88}, is that in this representation it is
straightforward to satisfy the diffeomorphism constraint, simply by
requiring that the wavefunctions be functions of loops invariant
under smooth deformations of the loops. Such functions are called knot
invariants in the mathematical literature. An unexpected connection
between knot theory and quantum gravity had been uncovered.

\section{Spin networks}

Unfortunately, the fact that the Wilson loop basis is over-complete
leaves an inconvenient imprint on the loop representation. For a
function of a given loop to be admissible as a wavefunction in the
loop representation, it has to satisfy the Mandelstam
identities. Therefore not any knot invariant will be suitable for a
wavefunction of the gravitational field. For years this problem
somewhat blocked progress in the construction of states.

However, it was noted by Rovelli and Smolin \cite{RoSmspin} three
years ago that there is a very natural way to label a basis of
independent Wilson loops. The construction is based on the notion of
spin network, introduced by Penrose \cite{Penrose} as a tool to
quantize gravity in an unrelated context in the sixties. The spin
networks of interest for quantum gravity, $\Gamma$, are graphs
embedded in three dimensions with three or higher valence
intersections. Each strand in the graph has associated with it an
element of the gauge group in a given representation, labelled by a
(half)integer $j$. The strands are ``tied up together'' at
intersections using invariant tensors in the group. The resulting
object, which we call ``Wilson net'' $W_\Gamma(A)$ is a generalization
to the spin network context of the trace of the holonomy of a single
loop. It is a gauge invariant object, and it can be shown that they
are free of Mandelstam identities. A way to believe this is to notice
that Mandelstam identities are an imprint left on holonomies due to
the use of a particular representation of the group in their definition
\cite{GaTr86}. Since Wilson nets contain all possible representations,
there are no identities present.

Having labelled an independent set of Wilson loops allowed to make
significant progress in other, apparently unrelated fronts. For many
years the issue of what sort of measures would one use to perform
integrals like (\ref{looptransf}) was an open question. These are
measures on an infinite-dimensional space with nonlinear constraints
(connections modulo gauge transformations). Having an independent
basis allows to ``do away with the nonlinearity'' and construct
examples of measures. This is a highly technical topic that was
pioneered by Ashtekar and Isham \cite{AsIs} and further developed by
Ashtekar, Baez, Lewandowski, Marolf, Mourao, Thiemann
\cite{AsLeMaMoTh}. To help tame the infinite dimensionality of the
space in questions, use is made of a special type of functions called
``cylindrical functions'' For this brief review, it will suffice to
say that with the measures introduced, the cylindrical functions are
orthogonal for different spin networks. That is, in the context of
these functions, the basis of spin networks not only is independent,
but in a sense it is an orthonormal basis. Moreover, these results can
be straightforwardly translated to the diffeomorphism invariant
context (where cylindrical functions are also defined). The statement
there would be that spin networks in different diffeomorphsim classes
are orthogonal.

\section{Physics at the kinematical level}

Having a space of functions that solves both the Gauss law and the
diffeomorphism constraint, endowed with an inner product preserved by
both constraints, it is tempting to see if one can compute things that
might have some physical interest. The quantities involved could be
considered a ``kinematical setting'' in terms of which to discuss
evolution. As we shall see, some interesting results arise that are
independent of the evolution chosen.

Some attractive statements can be made about areas and volumes in
terms of spin network states. If one is given a certain two
dimensional surface $S$ or a three dimensional one $\Sigma$, it is
possible to compute the area or volume in terms of the canonical
variables introduced. The expressions are,
\begin{eqnarray}
A &=& \int_S d^2 x \sqrt{\tilde{E}^a_i \tilde{E}^b_i n_a n_b},\\
V &=& \int_\Sigma d^3x \sqrt{\epsilon^{ijk} 
{\rlap{\lower1ex\hbox{$\sim$}}\epsilon{}}_{abc} 
\tilde{E}^a_i\tilde{E}^b_j\tilde{E}^c_k},
\end{eqnarray}
where $n_a$ is the normal of the surface. At first sight, the presence
of the square roots forecasts a regularization nightmare if one wishes
to promote these quantities to quantum operators. In fact, this is not
so. Because the operators have well defined properties under
diffeomorphisms and the correct density weights to be naturally
integrated, it turns out that their expressions are particularly
simple. For instance, the area of a surface evaluated on a spin
network state is proportional to $\sum_{j_i} \sqrt{j_i(j_i+1)}$, where
$j_i$ are the valences of the strands that pierce the surface. The
proportionality factor involves Planck's constant and the parameter
$\beta$. So one sees that in the basis of states considered, and with
the inner product we introduced in the last section, areas and volumes
are actually quantized and have a discrete spectrum
\cite{RoSmspin,AsLevo}. The elementary quantum of area involves the
parameter $\beta$ (also known as the Immirzi parameter). It appears
therefore that different choices of $\beta$ are associated with
different quantum theories with distinct predictions \cite{Im}.

The idea that areas are quantized and that the spin network strands
arise as ``elementary excitations'' embodies the concept of
``space-time foam'' in a concrete setting, and has found an attractive
use in attempts to explain black hole entropy. The idea is to view the
horizon of a black hole as a boundary of spacetime. The ``extra
degrees of freedom'' introduced by this boundary account for the
entropy of the black hole. This idea had been pursued in detail in the
$2+1$ dimensional context by Carlip \cite{Ca}. In the $3+1$ context,
using the formulation we are describing, the idea is that given a
certain area $A$ for the horizon, one could view the various
possibilities to obtain the given value of $A$ in terms of spin
networks as the ``degrees of freedom'' of the given area. This
requires a careful counting. Various countings have been proposed by
Smolin \cite{Sm}, Krasnov \cite{Kr}, Rovelli \cite{Ro}, and more
recently by Ashtekar, Baez, Corichi and Krasnov \cite{AsBaCoKr}. In
the latter work, a certain geometric condition fulfilled by classical
horizons (no outgoing radiation) is shown to imply the emergence of a
Chern--Simons theory on the boundary, which in turn allows a quite
precise counting. The end result is that the entropy, defined as
proportional to the logarithm of the number of states ends up being
proportional to the area of the surface, the constant of
proportionality involving the parameter $\beta$. The usual
Hawking--Bekenstein result of one-quarter the area can only be
achieved for a particular value of $\beta$. At the moment this issue
is still debated. If $\beta$ could be determined by independent means,
this would provide a check of the theory. It is encouraging that the
same value of $\beta$ is needed for different kinds of black holes.
Another striking property of the result is that it is largely
independent of the dynamics, which is only used at a classical level
in the arguments presented. There is ongoing work by various people on
all these issues.

\section{Dynamics}

As we mentioned before, one of the original motivations to use the
Ashtekar variables was that with the preferred choice $\beta=i$, the
Hamiltonian constraint became polynomial in the basic variables,
\begin{equation}
H = \epsilon^{ijk} \tilde{E}^a_i \tilde{E}^b_j F_{ab}^k.
\end{equation}
However, we noted that the resulting operator is a density of weight
$+2$, since it is quadratic in the momenta. This poses problems at the
time of regularizing the operator. There are no naturally defined
densities of weight two on a manifold. Therefore the end result of
most regularization attempts ends up being dependent on fiducial
backgrounds in terms of which one can construct the needed density
weights. The explicit appearance in the regularized operator of
artificial background structures implies very surely difficulties at
the time of enforcing the constraint algebra. For instance, in the
commutator of the Hamiltonian constraint and the diffeomorphism
constraint, external structures are not affected by the
diffeomorphisms generated by the constraint and therefore are not
covariant. This immediately leads to undesired anomalies.

Thiemann \cite{Th1-6} realized recently that one can represent the
apparently non-polynomial single-densitized Hamiltonian constraint,
\begin{equation}
H = {\epsilon^{ijk} \tilde{E}^a_i \tilde{E}^b_j \over \sqrt{det g}}F_{ab}^k.
\end{equation}
through the following classical identity,
\begin{equation}
{\epsilon^{ijk} \tilde{E}^a_i \tilde{E}^b_j \over \sqrt{det g}} =
\{A_c^i,V\}\tilde{\epsilon}^{abc}
\end{equation}
where $V$ is the volume we introduced in the previous section. The
resulting Hamiltonian therefore has the form,
\begin{equation}
H = \tilde{\epsilon}^{abc} {\rm Tr}( F_{ab}\,\{A_c,V\}),
\end{equation}
which with some care can be promoted very cleanly to a quantum
operator acting on diffeomorphism invariant cylindrical
functions. Thiemann has pursued this goal in a series of papers
\cite{Th1-6}, where he shows the operator is finite, well defined and
commutes with itself, therefore satisfying the correct constraint
algebra (if the space of functions were not diffeomorphism invariant,
the commutator of two Hamiltonians should be proportional to a
diffeomorphism, which automatically vanishes if the functions are
diffeomorphism invariant). In fact, the same procedure can be applied
to the other piece of the Hamiltonian constraint (if $\beta$ is real)
and to couplings to matter. Therefore Thiemann has constructed a
finite, consistent regularization of real general relativity coupled
to matter. This regularization implements the promise that
diffeomorphism invariance cures the divergences of field theories: the
resulting theory includes QCD and QED coupled to gravity with finite
Hamiltonians acting on a well defined space of diffeomorphism
invariant functions.

There is currently an active debate about the properties of Thiemann's
Hamiltonian. Most notably, Lewandowski and Marolf \cite{LeMa} have
noted that one can define another ``habitat'' where Thiemann's
Hamiltonian is well defined. This is a space of non--diffeomorphism
invariant functions that still share several properties of the usual
diffeo invariant cylindrical functions. They show that Thiemann's
regularization can be implemented in this space, but unfortunately,
the Hamiltonian that arises still commutes with itself, which is
inappropriate. This casts doubts about the regularization Thiemann
proposed, although no contradiction is proved about Thiemann's original
proposal. In collaboration with Gambini, Lewandowski and Marolf
\cite{GaLeMaPu} we have also proved that several modifications one
could propose to Thiemann's Hamiltonian do not cure the problem.

A separate line of attack is being pursued in terms of a different
space of functions. These functions are the knot invariants that come
from Chern--Simons theories. Some important mathematical hurdles were
cleared recently and progress is being made. But this topic will be
covered in detail in Gambini's lecture, so I will not describe it
here.

To summarize the situation for the dynamics, let us say that there is
a proposal (Thiemann's) that is a concrete, consistent, well defined
theory of canonical quantum gravity coupled to matter proposed by the
first time ever.  The proposal yields the correct dynamics in $2+1$
gravity as well. It is yet to be seen if it encompasses the correct
dynamics in $3+1$ dimensions. Some people suspect that the results of
Lewandowski and Marolf imply that it fails to do so, but there is no
clear proof of this yet. In the meantime, a separate proposal is being
worked out that might overcome these difficulties. Further studies
will determine if this is so.

\section{Conclusions}

I have attempted to give a quick review of several issues associated
with the current state of attempts to quantize gravity canonically. I
have left out many things. In particular I have not even referred to
many pieces of work that played crucial roles in the development of
the subject but that have become obsolete by the understanding they
themselves helped create. Among the current efforts I have failed to
cover, a great deal of activity is taking place these days trying to
find a ``covariant'' \cite{Bafoam,RoRe,MaSm} 
formulation of the theory using path
integrals. One of the rationales for this is that these kinds of
formulations might offer a fresh look on what kind of modifications to
make to Thiemann's Hamiltonian to recover the correct dynamics. At the
moment this work is largely exploratory.

The possibility to have for the first time ever a consistent, finite,
well defined theory of quantum gravity coupled to matter should not be
under-stressed. The payoff is big: not only is gravity quantized, but
all divergences of quantum field theory disappear. It is quite natural
and understandable that the first theories we might generate end up
being later considered with the hindsight of time ``trivial'' or
``wrong". However, it is obviously important then to pursue the
current theories further to see if they corresponds to the quantum
gravity we expect to see in nature or to inapplicable mathematical
elaborations. This is what one expects from science.

\section{Acknowledgements}

I wish to thank the organizers for the invitation and financial
support.  This work was supported in part by grants NSF-INT-9406269,
NSF-PHY-9423950, research funds of the Pennsylvania State University,
the Eberly Family research fund at PSU and PSU's Office for Minority
Faculty development. I also acknowledge support of the Alfred P. Sloan
foundation through a fellowship.

\end{document}